\documentclass[11pt,twoside,twocolumn]{article}

\usepackage{amsfonts,amsmath,amssymb,bm,abstract,balance,graphicx}

\title{\textbf{Neutrino flavor oscillations in rotating matter}}

\author{Maxim Dvornikov%\thanks{E-mail: maxim.dvornikov@usm.cl}
\\
\small{Institute of Physics, University of S\~{a}o Paulo,} \\
\small{CP 66318, CEP 05315-970 S\~{a}o Paulo, SP, Brazil; and} \\
\small{Pushkov Institute of Terrestrial Magnetism, Ionosphere} \\
\small{and Radiowave Propagation (IZMIRAN),} \\
\small{142190, Troitsk, Moscow region, Russia} \\
\small{E-mail: maximd@if.usp.br}}

\date{}

\begin{document}

\twocolumn[\maketitle
\begin{onecolabstract}
We study the evolution of the neutrinos system in
rotating matter. Neutrinos are supposed to be mixed massive
particles interacting with background fermions by means of the
electroweak forces. First we find the solutions of wave equations
for the neutrino mass eigenstates in matter. Then we study the
behavior of neutrino flavor eigenstates in background matter. The
problems of neutrino bound states and neutrino flavor oscillations
are discussed. We also derive the analog of the quantum mechanical
evolution equation for the system of two flavor neutrinos in
rotating matter and analyze its solution for the particular
initial condition for neutrino flavor eigenstates.
\\
\textbf{Keywords}: exact solutions of wave equations, neutrino oscillations, rotating matter
\\
\end{onecolabstract}]

\section{Introduction}%\label{sec:Introduction}

Nowadays it is acknowledged that neutrinos play a significant role
in the evolution of massive stars at the ultimate stages of their
life. When the mass of a star is about $8-9$ or $40-60$ solar
masses, such a star can explode as a supernova with the emission
of great quantity of neutrinos carrying away almost $99\%$ of the
initial gravitational energy of a star~\cite{GiuKim07}.

Although neutrinos interact rather weakly with background matter,
these particles are the key component in the dynamics of a
supernova explosion. After the supernova explosion the core a
massive star is converted into a compact dense object, a neutron
star. Neutrinos are supposed to give the contribution to the
subsequent evolution of a neutron star, i.e. causing its
cooling~\cite{Yak01}. Besides the direct influence to the
supernova explosion process neutrinos can also affect various
macroscopic characteristics of a neutron star. For example, great
peculiar velocities of pulsars can be explained by the asymmetric
neutrino emission~\cite{KusSer96}. It is also supposed that the
emission of neutrinos can cause the spin-down of a rotating
neutron star~\cite{sd}.

The fact that neutrinos are massive particles has many important
consequences. Unlike photons that immediately escape the region
where they are created, neutrinos with relatively small initial
energies can form bound states inside or in the vicinity of
various astrophysical objects. The most proper candidates for the
formation of such non-trivial states are relic neutrinos. The
possibility of gravitational clustering of relic neutrinos was
studied in Ref.~\cite{RinWon04}. It was discussed in
Ref.~\cite{superf} that both Dirac and Majorana neutrinos can also
create a superfluid condensate due to the Higgs boson
interactions. We considered the situation of gravitational
trapping of neutrinos by a massive black hole in
Ref.~\cite{Dvo06}. The neutrino trapping in both curved space-time
and rotating matter was studied in Ref.~\cite{Tha78}.

There is also a possibility that neutrinos emitted in a neutron
star form bound orbits inside the star due to their collective
interactions with neutron matter. This issue was studied in
Ref.~\cite{nuinsideNS}. Neutrino trapping inside a rotating
neutron star was discussed in Ref.~\cite{StuROT}. This effect
results from the neutrino electroweak interaction with
inhomogeneously moving matter.

Recently we developed an approach for the description of neutrino
flavor and spin-flavor oscillations in various external
fields~\cite{approach}. The problem is formulated in terms of wave
quantum mechanics and involves exact solutions to wave equations
for a neutrino in an external field. In the present work we study
the evolution of massive mixed neutrinos in inhomogeneously moving
matter. In particular, we apply the treatment to the emission of
neutrinos in a rotating neutron star.

In Sec.~\ref{FO} we give the general formulation of two neutrino
flavors interacting with rotating matter of the type found in
neutron stars. In Secs.~\ref{APP} and~\ref{DirEqMASS} we find the
solutions for the Dirac equation for a neutrino interacting with
moving matter for massless and massive particles respectively. We
also compare our solutions with the previously found
ones~\cite{StuROT,BalPopStu09}. In Sec.~\ref{BO} we discuss the
possibility for low energy neutrinos to form bound orbits inside a
neutron star. Neutrino flavor oscillations in rotating matter are
discussed in Sec.~\ref{LO}. In Appendix~\ref{INICOND} we state the
solution of the wave equation for a neutrino in vacuum in
cylindrical coordinates. Finally in Sec.~\ref{CONCL} we summarize
our results.

\section{General formulation\label{FO}}

Let us first formulate  the evolution of two neutrino flavor
eigenstates, $\nu_\lambda$, $\lambda = \alpha,\beta$, interacting
with moving matter due to the exchange of the electroweak $Z$ and
$W^{\pm{}}$ bosons. Phenomenologically this interaction with
matter can be implemented by means of a set of neutrino wave
equations with the external fields $f^\mu_\lambda$ \cite{MohPal04}
shown below. In the flavor basis, neutrinos also have a
non-diagonal mass matrix $(m_{\lambda\lambda'})$. The Lagrangian
for this system is then:
\begin{align}\label{Lagrnu}
  \mathcal{L} = &
  \sum_{\lambda=\alpha,\beta}
  \bar{\nu}_\lambda
  (\mathrm{i}\gamma^\mu\partial_\mu - f^\mu_\lambda\gamma_\mu P_\mathrm{L})
  \nu_\lambda
  \notag
  \\
  & -
  \sum_{\lambda\lambda'=\alpha,\beta}
  m_{\lambda\lambda'} \bar{\nu}_\lambda \nu_{\lambda'},
  \notag
  \\
  & P_\mathrm{L} = \frac{1}{2}(1-\gamma^5).
\end{align}
In our case of interest, the flavor $\alpha$ will be either $\mu$
or $\tau$, and the flavor $\beta$ will be $e$.

The form of the currents $f^\mu_\lambda$ are determined by the
neutrino interactions with the medium. In the case of a neutron
star the medium consists of electrons, protons, and neutrons, with
number densities $n_e$, $n_p$ and $n_n$, respectively, and $n_e =
n_p$, corresponding to electrically neutral matter. Therefore, for
the standard model neutrino flavors $\alpha$ (either $\mu$ or
$\tau$) and $\beta = e$, the corresponding external fields have
the following expressions~\cite{DvoStu02JHEP}:
\begin{equation}\label{falphabeta}
  f_\alpha^\mu = -\frac{G_\mathrm{F}}{\sqrt{2}}j_n^\mu,
  \quad
  f_\beta^\mu = \frac{G_\mathrm{F}}{\sqrt{2}}(2 j_e^\mu-j_n^\mu),
\end{equation}
where $G_\mathrm{F}$ is the Fermi constant and
\begin{equation}\label{je}
  j_e^\mu = (n_e, n_{e}\mathbf{v}),
  \quad
  j_n^\mu = (n_n, n_{n}\mathbf{v}),
\end{equation}
are the hydrodynamical currents of each of the %%@
background fermion species. We also assume that all background
fermions rotate as a rigid body, i.e., moving with same velocity
$\mathbf{v}$.

To study the evolution of the system~\eqref{Lagrnu} we diagonalize
the mass matrix $(m_{\lambda\lambda'})$ by introducing the set of
the neutrino mass eigenstates $\psi_a$, $a=1,2$, with help of the
matrix transformation:
\begin{align}\label{matrtrans}
  \nu_{\lambda} = & \sum_{a=1,2}U_{\lambda a}\psi_a,
  \notag
  \\
  {(U_{\lambda a})} = &
   \begin{pmatrix}
    U_{\alpha 1} & U_{\alpha 2} \\
    U_{e 1} & U_{e 2} \
  \end{pmatrix}
  =
  \begin{pmatrix}
    \cos \theta & -\sin \theta \\
    \sin \theta & \cos \theta \
  \end{pmatrix}.
\end{align}
Here $\theta$ is the vacuum mixing angle, where $\theta=0$ means
$\nu_1 = \nu_\alpha$ and $\nu_2 = \nu_e$. After diagonalization
the Lagrangian~\eqref{Lagrnu} reads
\begin{align}\label{Lagrpsi}
  \mathcal{L} = &
  \sum_{a=1,2}
  \bar{\psi}_a (\mathrm{i}\gamma^\mu\partial_\mu-m_a) \psi_a 
  \notag
  \\
  & -
  \sum_{a,b=1,2}
  g^\mu_{ab} \bar{\psi}_a \gamma_\mu P_\mathrm{L} \psi_{b},
\end{align}
where the matter interaction term contains the 2$\times$2 matrix
$(g^\mu_{ab})$ in the mass eigenstate basis
\begin{multline}\label{gab}
  (g^\mu_{ab}) =
  \frac{G_\mathrm{F}}{\sqrt{2}}
  \\
  %\notag
  \times
  \begin{pmatrix}
    [2 j_e^\mu \sin^2\theta-j_n^\mu] & j_e^\mu \sin 2\theta \\
    j_e^\mu \sin 2\theta & [2 j_e^\mu \cos^2\theta-j_n^\mu] \
  \end{pmatrix}.
\end{multline}

The Dirac equation for the neutrino mass eigenstates, obtained
from Eq.~\eqref{Lagrpsi}, has then the form,
\begin{multline}\label{DireqpsiMP}
  (\mathrm{i}\gamma^\mu\partial_\mu-m_a - g^\mu_{aa} \gamma_\mu
  P_\mathrm{L})\ \psi_a
  \\
  -
  g^\mu_{ab} \gamma_\mu P_\mathrm{L}\ \psi_{b}=0 ,
  \quad a \neq b,
\end{multline}
where the last term corresponds to the off-diagonal elements of
Eq.~\eqref{gab}, so it is an interaction with the medium that
mixes the different mass eigenstates.

We solve Eq.~\eqref{DireqpsiMP} treating the last term (the mixing
of neutrino types) as a perturbation, so that at zeroth order the
neutrino types are decoupled. Also, since the typical energy of a
neutrino emitted in a neutron star is $\sim
10\thinspace\text{MeV}$ whereas the neutrino masses do not exceed
a few eV, the neutrinos are ultrarelativistic and we can treat the
masses $m_a$ also as perturbations.

Now, for a rigid rotating medium, the interaction depends on a
velocity $\mathbf{v(r)= \Omega\times r}$, where $\mathbf{r}$ is
the radius vector from the star center and $\mathbf{\Omega}$ is
the angular velocity of the star. Therefore we define the positive
potentials as:
\begin{multline}\label{lambda12}
  V_a = - g^0_{aa} 
  \\
  =
  \begin{cases}
    G_\mathrm{F}
    (n_n-2 n_e \sin^2\theta)/\sqrt{2}, & \text{for $a=1$}, \\
    G_\mathrm{F}
    (n_n-2 n_e \cos^2\theta)/\sqrt{2}, &
    \text{for $a=2$}.
  \end{cases}
\end{multline}
Note that $V_a>0$ in Eq.~\eqref{lambda12} since $n_n\gg n_e$ in a
neutron star and $g^0_{aa}<0$ ($a=1,2$).

In order to proceed with the description of the evolution of the
system~\eqref{DireqpsiMP} we should have the energy levels and
wave fuctions of the mass eigenstates $\psi_a$ which correspond to
the wave equation~\eqref{DireqpsiMP} at the absence of the mixing
term $g^\mu_{ab}$. These quantities will be found in
Secs.~\ref{APP} and~\ref{DirEqMASS}.

\section{Solution of the wave equation for a neutrino in rotating
matter in the limit of zero mass\label{APP}}

In this section we find the solution of Eq.~\eqref{DireqpsiMP} at
the absence of the mixing between different mass eigenstates due
to the interaction with matter, i.e. we put the coefficient
$g_{12}^\mu$ to zero.

This case corresponds to a single unmixed neutrino interacting
with an external axial-vector field. The wave
equation~\eqref{DireqpsiMP} is transformed to the
form~\cite{DvoStu02JHEP,LobStu01},
\begin{equation}\label{Direqpsi}
  (\mathrm{i} \gamma^\mu \partial_\mu -
  m - g^\mu \gamma_\mu P_\mathrm{L})\psi = 0.
\end{equation}
Here we are interested in the case of ultrarelativistic neutrinos,
i.e when the mass in Eq.~\eqref{Direqpsi} is much smaller than the
energy.

The equations of motion for the left-handed $\eta$ and
right-handed $\xi$ chiral components of the spinor
$\psi^\mathrm{T}=(\xi,\eta)$ decouple in the $m=0$ limit. The mass
contribution can be included in perturbation theory (see
Sec.~\ref{DirEqMASS}). Using the chiral basis for the $\gamma^\mu$
matrices in the convention of Ref.~\cite{ItzZub80},
\begin{gather}\label{chiralbasis}
  \gamma^0 =
  \left(
    \begin{array}{ll}
      0 & -1 \\
      -1 & 0 \
    \end{array}
  \right),
  \quad
  \bm{\gamma} =
  \left(
    \begin{array}{ll}
      0 & \bm{\sigma} \\
      -\bm{\sigma} & 0 \
    \end{array}
  \right),
  \notag
  \\
  \gamma^5 =
  \left(
    \begin{array}{rr}
      1 & 0 \\
      0 & -1 \
    \end{array}
  \right),
\end{gather}
the Dirac equation for the left-handed component $\eta$ of the
neutrino is
\begin{equation}\label{dotetaeq}
  \mathrm{i}\dot{\eta} =
  \{
  \mathrm{i}\bm{\sigma}\cdot\mathbf{\nabla} +
  \bar\sigma_\mu g^\mu
  \}\eta,
\end{equation}
where the matter interaction term is $\bar\sigma_\mu g^\mu = g_0
+\bm{\sigma}\cdot\mathbf{g}$.

The external field $g^\mu$ with non-zero spatial components
corresponds to the external electroweak field due to the moving
background matter. In analogy to Eq.~\eqref{gab} we obtain for the
three-vector part, $\mathbf{g} = g_0 \mathbf{v}$. We want to study
the neutrino states inside a neutron star rotating with the
angular velocity $\bm{\Omega}$ directed along the $z$ axis,
$\bm{\Omega}=\Omega\mathbf{e}_z$. As we mentioned in
Sec.~\ref{FO}, since $g_0 < 0$ in the case of a neutron star, we
redefine this potential as $V = -g_0$. Accordingly,
\begin{equation}\label{lambda}
  \mathbf{g}=V \Omega(y\mathbf{e}_x - x\mathbf{e}_y).
\end{equation}
It is natural to use cylindrical coordinates $(r,\phi,z)$ to solve
Eq.~\eqref{dotetaeq} with
\begin{align}\label{spsg}
  \mathrm{i} \bm{\sigma}\cdot\mathbf{\nabla} = &
  \mathrm{i}
  \begin{pmatrix}
    \partial_z & e^{-\mathrm{i}\phi}[\partial_r-(\mathrm{i}/r)\partial_\phi] \\
    e^{\mathrm{i}\phi}[\partial_r+(\mathrm{i}/r)\partial_\phi] & -\partial_z \
  \end{pmatrix},
  \notag
  \\
  \bm{\sigma}\cdot\mathbf{g}= &
  V \Omega r
  \begin{pmatrix}
    0 & \mathrm{i} e^{-\mathrm{i}\phi} \\
    -\mathrm{i} e^{\mathrm{i}\phi} & 0 \
  \end{pmatrix}.
\end{align}
Taking into account that Eqs.~\eqref{dotetaeq} and~\eqref{spsg} do
not depend on $z$, we look for a stationary solution of the form
$\eta \sim u(r,\phi) e^{-\mathrm{i}(E t - p_z z)}$. Using the
analysis of Ref.~\cite{SokTer74} we can further separate the two
coordinates $r$ and $\phi$ in the two-component spinor $u$ using
auxiliary functions $F_{1,2}(\rho)$ as
\begin{equation}\label{umassless}
  u(r,\phi) =\frac{1}{\sqrt{2\pi}}
  \begin{pmatrix}
    e^{\mathrm{i} (l-1) \phi} F_1(\rho)\\
    e^{\mathrm{i} l \phi} F_2(\rho)
  \end{pmatrix},
\end{equation}
where $\rho = V\Omega r^2$, and $l$ is an integer so that the
function of $\phi$ is single-valued. Since the system is invariant
under rotations around the $Z$ axis, $u$ has to definite value of
$J_z$ (total angular momentum), which in our notation is equal to
$l-1/2$.

The functions $F_{1,2}$ satisfy the coupled equations:
\begin{align}\label{RF}
  \mathrm{i} \sqrt{V\Omega\rho}
  &
  \left(
    2 \partial_\rho - 1 - \frac{l-1}{\rho}
  \right) F_1 = \mathrm{i}R_1F_1 
  \notag
  \\
  & = (E-p_z+V) F_2,
  \notag
  \\
  \mathrm{i} \sqrt{V\Omega\rho}
  &
  \left(
    2 \partial_\rho + 1 + \frac{l}{\rho}
  \right) F_2 = \mathrm{i}R_2F_2
  \notag
  \\
  & = (E+p_z+V) F_1,
\end{align}
which can be separated into two independent second order
differential equations:
\begin{align}
  \label{F1eq}
  \bigg\{
    \rho \frac{\mathrm{d}^2}{\mathrm{d}\rho^2}
   +
    \frac{\mathrm{d}}{\mathrm{d}\rho}
    + &
    \kappa - \frac{l}{2} 
    \notag
    \\
    & - \frac{\rho}{4} - \frac{(l-1)^2}{4\rho}
  \bigg\} F_1 = 0,
  \\
  \label{F2eq}
  \bigg\{
    \rho \frac{\mathrm{d}^2}{\mathrm{d}\rho^2}
    +
    \frac{\mathrm{d}}{\mathrm{d}\rho}
    + &
    \kappa - \frac{l-1}{2} 
    \notag
    \\
    & - \frac{\rho}{4} - \frac{l^2}{4\rho}
  \bigg\} F_2 = 0,
\end{align}
where the parameter $\kappa$ is related to the energy in the form,
\begin{equation}\label{kappa}
  \kappa = \frac{(E+V)^2-p_z^2}{4V\Omega}.
\end{equation}
Let us just solve Eq.~\eqref{F1eq} for the function $F_1$.
Eq.~\eqref{F2eq} can be solved for $F_2$ by analogy. Expressing
$F_1(\rho) = e^{-\rho/2}\rho^{(l-1)/2}u(\rho)$, the new function
$u(\rho)$ obeys an \emph{associated Laguerre equation},
\begin{equation}\label{ueq}
  \rho u'' + (l-\rho)u' + (\kappa-l) u = 0,
\end{equation}
whose solutions are the associated Laguerre polynomials $u(\rho)
\sim Q_s^{l-1}(\rho)$. Here $s = \kappa-l$ is the radial quantum
number, and $\kappa$, defined in Eq.~\eqref{kappa}, is related to
the neutrino energy. The function $F_1$ is then a \emph{Laguerre
function}, $F_1(\rho) = I_{\kappa-1,s}(\rho)$. The Laguerre
functions $I_{{\rm n},s}(\rho)$ are defined in terms of the
associated Laguerre polynomials $Q_{s} ^{l} (\rho)$ (where ${\rm
n}= s+l$) as
\begin{align}
  I_{{s+l},s}(\rho) = & \frac{1}{\sqrt{{(s+l)}!s!}}
  e^{-\rho/2}\rho^{l/2}Q_s^{l}(\rho),
 \notag
 \\
 \quad Q_s^l (\rho) = & e^\rho \rho^{-l}
  \frac{\mathrm{d}^s}{\mathrm{d}\rho^s}(\rho^{s+l} e^{-\rho}) .
\end{align}
Another common definition of the associated Laguerre polynomials
is $L_s ^l (\rho) = Q_s ^l (\rho)/ s!$ The Laguerre polynomials
satisfy the recursive relation,
\begin{equation}\label{recurr}
 Q_s^{l-1}(\rho) =Q_s^l (\rho) - s Q_{s-1}^{l}(\rho).
\end{equation}
We also mention that $I_{{\rm n},s}(\rho)$ and $Q_s ^l (\rho)$
satisfy the integral relations,
\begin{align}\label{integrals}
  \int_0^\infty I_{{\rm n},s}(\rho) I_{{\rm n}-1,s}(\rho)\sqrt{\rho}\ \mathrm{d}\rho & =
  \sqrt{{\rm n}},
  \notag
  \\
  \int_0^\infty e^{-\rho} \rho^l Q_s^l(\rho) Q_{s'}^l(\rho)\ \mathrm{d}\rho & =
  \delta_{ss'} s! (l+s)!
\end{align}

In order to have normalizable functions at the origin, $l$ must be
a non-negative integer. Also, the solution diverges at large
radii, unless $s$ is a non-negative integer. Therefore, in order
to have well-behaved solutions, the smaller values of $s$ and
$\kappa$ must be integers, in which case the lower values of
energy, $E = -V \pm \sqrt{4 V\Omega \kappa + p_z^2}$, are
discrete:
\begin{equation}\label{enlev}
  \kappa\to {\rm n} , \quad
  E\to E_{\rm n} = -V \pm \sqrt{4 V\Omega\ {\rm n} + p_z^2}.
\end{equation}
On the other hand, for large enough energies the divergent
behavior of the solution falls outside the star radius, in which
case $\kappa$ has no restrictions and it becomes a continuous
variable.

Notice that the energy levels in Eq.~\eqref{enlev} are different
from the analogous expressions obtained in Ref.~\cite{StuROT}, $E
= -V \pm \sqrt{2 V\Omega n + p_z^2}$. It was claimed in
Ref.~\cite{StuROT} that there is an analogy between the charged
particle dynamics in an electromagnetic field and the neutrino
motion in matter. However this analogy is just superficial. An
electromagnetic field is gauge invariant. Therefore when we study
the motion of an electron in an external magnetic field
$\mathbf{B}=(0,0,B)$, we can choose any gauge. In a gauge that
keeps the cylindrical symmetry explicit, the vector potential
$\mathbf{A}=(-yB/2,xB/2,0)$ is required. The analog of this gauge
is adopted in the present work. If the electron motion is studied
in cartesian coordinates, the Landau gauge is more convenient,
$\mathbf{A}=(0,xB,0)$. This kind of gauge was used in
Ref.~\cite{StuROT}. As shown in Ref.~\cite{SokTer74}, in the
electromagnetic problem both gauges must give the same energy
spectrum for the electron. However for the motion of a neutrino in
rotating matter, the situation is different: there is no gauge
freedom in this case. If one uses the analog of the Landau gauge,
as in Ref.~\cite{StuROT}, one underestimates the matter
contribution to the dispersion relation.

Finally, to derive the lower component of the neutrino spinor $u$,
we just need to use the identities (see Ref.~\cite{SokTer74}),
\begin{align}
  R_1 I_{{\rm n}-1,s}(\rho) = &
  -\sqrt{4 V\Omega {\rm n}} I_{{\rm n},s}(\rho),
  \notag
  \\
  R_2 I_{{\rm n},s}(\rho) = &
  \sqrt{4 V\Omega {\rm n}} I_{{\rm n}-1,s}(\rho),
\end{align}
to find the function $F_2(\rho)$. Thus we get the complete
two-component spinor in the form
\begin{equation}\label{etaabs}
  u(r,\phi) =
  \frac{\sqrt{2V\Omega}}{\sqrt{2\pi}}
  \begin{pmatrix}
    C_1 I_{{\rm n}-1,s}(\rho) e^{\mathrm{i}(l-1)\phi} \\
    \mathrm{i} C_2 I_{{\rm n},s}(\rho) e^{\mathrm{i}l\phi} \
  \end{pmatrix}.
\end{equation}
The coefficients $C_{1,2}$ are related to each other due to
Eq.~\eqref{RF} as
\begin{equation}\label{C1C2}
  \sqrt{4V\Omega {\rm n}}\ C_1+(E+V-p_z)C_2=0,
\end{equation}
and their norm can be chosen to satisfy $C_1^2+C_2^2=1$.

Let us now discuss the limit of small angular velocities to
establish the connection with the non-rotating case. For
simplicity we study the case $p_z = 0$ which is examined in
Sec.~\ref{FO}. The limit $\Omega \to 0$ should be taken together
with ${\rm n} \to \infty$, so that $\Omega\cdot {\rm n} =
\text{constant}$. Using the identity~\cite{SokTer74p282},
\begin{equation}\label{IJrelgen}
  \lim_{{\rm n} \to \infty}
  I_{{\rm n},{\rm n}-l}([p_\perp r]^2/4{\rm n}) =
  J_l(p_\perp r),
\end{equation}
we reproduce the neutrino wave functions in
vacuum~\eqref{eta0cylvac}. Therefore one can identify $\sqrt{4
V\Omega {\rm n}}$ with the neutrino momentum in the equatorial
plane, $p_\perp$, in Eq.~\eqref{enlev}.

\section{Approximate solution of the wave equation
for a massive neutrino in rotating matter\label{DirEqMASS}}

In this section we will study the effect of the rotation of matter
on the single massive neutrino without mixing, analogously to the
treatment of Sec.~\ref{APP}. However, now we will take into
account the contribution of the neutrino mass to the energy
levels~\eqref{enlev} using the perturbation theory.

For a massive particle one should take into account both spinors
$\xi$ and $\eta$ in Eq.~\eqref{Direqpsi}. The coupled wave
equations for these spinors have the following form:
\begin{align}\label{eqetaxim}
  \mathrm{i} \dot{\xi}= &
  (\bm{\sigma} \cdot \mathbf{p})\xi - m \eta,
  \notag
  \\
  \mathrm{i} \dot{\eta}= &
  -(\bm{\sigma} \cdot \mathbf{p})\eta - m \xi +
  [g_0 + (\bm{\sigma} \cdot \mathbf{g})] \eta,
\end{align}
where the vector $\mathbf{g}$ is defined in Eq.~\eqref{lambda}.

Looking for the stationary solutions of Eq.~\eqref{eqetaxim}, $\xi
\sim e^{-\mathrm{i}Et}$ and $\eta \sim e^{-\mathrm{i}Et}$, and
excluding spinor $\xi$ from Eq.~\eqref{eqetaxim} we get the only
differential equation for the spinor $\eta$,
\begin{multline}\label{etam}
  [E^2 - m^2 - \mathbf{p}^2 + VE - E(\bm{\sigma} \cdot \mathbf{g}) 
  \\
  -
  V(\bm{\sigma} \cdot \mathbf{p}) 
  % \notag
  % \\
  +
  (\bm{\sigma} \cdot \mathbf{p})(\bm{\sigma} \cdot \mathbf{g})] \eta = 0.
\end{multline}
Note that one should take into account the non-commutativity of
the operator $\mathbf{p}$ and vector $\mathbf{g}$ since the latter
depends on the spatial coordinates.

As in Sec.~\ref{APP} we will use cylindrical coordinates $(r,\phi,
z)$ to analyze Eq.~\eqref{etam}. It is convenient to rewrite this
equation for each of the components of the spinor
$\eta^\mathrm{T}=(\eta_1,\eta_2)$,
\begin{align}\label{etamcomp}
  \bigg\{
    E^2 - & m^2  + VE +
    \partial_r^2 + \frac{1}{r}\partial_r + \frac{\partial_\phi^2}{r^2} + \partial_z^2
    \notag
    \\ & 
    +
    \mathrm{i}V\Omega\partial_\phi
    +
    \mathrm{i}V\partial_z 
    + 
    V\Omega (r\partial_r + 2)
  \bigg\} \eta_1
  \notag
  \\
  & =
  - e^{-\mathrm{i}\phi} V
  \bigg\{
    \mathrm{i}
    \left[
      \partial_r - \frac{\mathrm{i}}{r}\partial_\phi
    \right] 
    \notag
    \\
    & -
    \Omega r (\mathrm{i} E - \partial_z)
  \bigg\}\eta_2,
  \notag
  \\
  \bigg\{
    E^2 - & m^2  + VE +
    \partial_r^2 + \frac{1}{r}\partial_r + \frac{\partial_\phi^2}{r^2} + \partial_z^2
    \notag
    \\ & 
    +
    \mathrm{i}V\Omega\partial_\phi
    -
    \mathrm{i}V\partial_z - V\Omega (r\partial_r + 2)
  \bigg\}\eta_2
  \notag
  \\
  & =
  - e^{\mathrm{i}\phi} V
  \bigg\{
    \mathrm{i}
    \left[
      \partial_r + \frac{\mathrm{i}}{r}\partial_\phi
    \right] 
    \notag
    \\ &
    +
    \Omega r (\mathrm{i} E + \partial_z)
  \bigg\}\eta_1.
\end{align}
We look for the solution of Eq.~\eqref{etamcomp} in the following
form:
\begin{align}
  \eta = & e^{\mathrm{i}p_z z} u(r,\phi),
  \notag
  \\
  u(r,\phi) = &
  \frac{1}{\sqrt{2\pi}}
  \begin{pmatrix}
    e^{\mathrm{i} (l-1) \phi} F_1(r) \\
    e^{\mathrm{i} l \phi} F_2(r) \
  \end{pmatrix},
\end{align}
where $F_{1,2}(r)$ are the new unknown functions [see
Eq.~\eqref{umassless}]. From Eq.~\eqref{etamcomp} we derive the
equations for the functions $F_{1,2}(r)$,
\begin{align}\label{F12mcomp}
  \bigg\{
    E^2 - & m^2 - p_z^2 + VE +
    \partial_r^2 + \frac{1}{r}\partial_r     
    - 
    \frac{(l-1)^2}{r^2} 
    \notag
    \\ &
    -
    V\Omega(l-1)
    -
    V p_z + V\Omega (r\partial_r + 2)
  \bigg\} F_1
  \notag
  \\
  & =
  - \mathrm{i}V
  \bigg\{
    \left[
      \partial_r + \frac{l}{r}
    \right] 
    \notag
    \\ &    
    -
    \Omega r (E - p_z)
  \bigg\}F_2,
  \notag
  \\
  \bigg\{
    E^2 - & m^2 - p_z^2 + VE +
    \partial_r^2 + \frac{1}{r}\partial_r - \frac{l^2}{r^2} 
    \notag
    \\ &    
    -
    V\Omega l
    +
    V p_z - V\Omega (r\partial_r + 2)
  \bigg\}F_2
  \notag
  \\
  & =
  - \mathrm{i} V
  \bigg\{
    \left[
      \partial_r - \frac{(l-1)}{r}
    \right] 
    \notag
    \\ &    
    +
    \Omega r (E + p_z)
  \bigg\}F_1.
\end{align}

Introducing the dimensionless variable $\rho = V\Omega r^2$, as in
Sec.~\ref{APP}, and using the properties of the operators
$R_{1,2}$, defined in Eq.~\eqref{RF},
\begin{align}
  R_1 R_2 = & 4 V\Omega
  %\notag
  % \\ &
  % \times
  \bigg(
    \rho\partial_\rho + \partial_\rho
    \notag
    \\ &  
    -
    \frac{l-1}{2} - \frac{\rho}{4} - \frac{l^2}{4\rho}
  \bigg),
  \notag
  \\
  R_2 R_1 = & 4 V\Omega
  %\notag
  %\\ &
  %\times
  \bigg(
    \rho\partial_\rho + \partial_\rho
    \notag
    \\ & 
    -
    \frac{l}{2} - \frac{\rho}{4} - \frac{(l-1)^2}{4\rho}
  \bigg),
\end{align}
we rewrite Eq.~\eqref{F12mcomp} in the following form:
\begin{align}\label{F12R12}
  \big\{
    E^2 - & m^2 - p_z^2 + VE - V p_z + 3V\Omega 
    \notag
    \\ & 
    +
    R_2 R_1 + \sqrt{V\Omega\rho}R_2
  \big\} F_1
  \notag
  \\
  & =
  - \mathrm{i}
  \big\{
    V R_2 - \sqrt{V\Omega\rho}
    \notag
    \\ &
    \times
    (V+E-p_z)
  \big\}F_2,
  \notag
  \displaybreak[1]
  \\
  \big\{
    E^2 - & m^2 - p_z^2 + VE - V p_z - 3V\Omega 
    \notag
    \\ & 
    +
    R_1 R_2 - \sqrt{V\Omega\rho}R_1
  \big\} F_2
  \notag
  \\
  & =
  - \mathrm{i}
  \big\{
    V R_1 + \sqrt{V\Omega\rho}
    \notag
    \\ &
    \times
    (V+E+p_z)
  \big\}F_1.
\end{align}

To study the contribution of the neutrino mass to the neutrino
energy spectrum~\eqref{enlev} we discuss the situation of the
neutrino bound states and take into account neutrino mass with
help of the perturbation theory. It means that in
Eq.~\eqref{F12R12} we can use the expressions for the wave
functions $F_{1,2}$ presented in Eq.~\eqref{etaabs} which
correspond to the massless neutrino: $F_1(\rho) = C_1
I_{\mathrm{n}-1,s}(\rho)$ and $F_2(\rho) = \mathrm{i} C_2
I_{\mathrm{n},s}(\rho)$. For this kind of wave functions we get
from Eq.~\eqref{F12R12} the following relations:
\begin{align}\label{C1C2rel}
  \big\{
    C_1 & [E^2 - m^2 - p_z^2 
    \notag
    \\ &
    + 
    V(E - p_z + 4\Omega -
    4\Omega\mathrm{n})]
    \notag
    \\ &
    -
    C_2 V \sqrt{4V\Omega\mathrm{n}}
  \big\} I_{\mathrm{n}-1,s}(\rho)
  \notag
  \\
  & +
  \sqrt{V\Omega\rho}(V+E-p_z) C_2 I_{\mathrm{n},s}(\rho)
  \notag
  \\
  & +
  2V\Omega\sqrt{\rho(\mathrm{n}-1)} C_1
  I_{\mathrm{n}-2,s}(\rho)=0,
  \notag
  \\
  - \big\{
    C_2 & [E^2 - m^2 - p_z^2 
    \notag
    \\ &    
    + 
    V(E + p_z - 4\Omega -
    4\Omega\mathrm{n})]
    \notag
    \\ & 
    -   
    C_1 V \sqrt{4V\Omega\mathrm{n}} 
  \big\} I_{\mathrm{n},s}(\rho)
  \notag
  \\
  \notag
  & -
  \sqrt{V\Omega\rho}(V+E+p_z) C_1 I_{\mathrm{n}-1,s}(\rho)
  \notag
  \\ &
  -
  2V\Omega\sqrt{\rho(\mathrm{n}+1)} C_2
  I_{\mathrm{n}+1,s}(\rho)=0.
\end{align}
To obtain Eq.~\eqref{C1C2rel} we use the known properties of the
Laguerre functions,
\begin{align}
  I_{\mathrm{n}+1,s}(\rho) = &
  \sqrt{\frac{\rho}{\mathrm{n}+1}}
  \notag
  \\ &
  \times
  \left[
    \frac{\rho+\mathrm{n}-s}{2\rho}I_{\mathrm{n},s}(\rho)-
    I_{\mathrm{n},s}'(\rho)
  \right],
  \notag
  \displaybreak[1]
  \\
  I_{\mathrm{n}-2,s}(\rho) = &
  \sqrt{\frac{\rho}{\mathrm{n}-2}}
  \notag
  \\ &
  \times
  \bigg[
    \frac{\rho+\mathrm{n}-s-1}{2\rho}I_{\mathrm{n}-1,s}(\rho)
    \notag
    \\ &
    +
    I_{\mathrm{n}-1,s}'(\rho)
  \bigg],
\end{align}
which can be found in Ref.~\cite{Bor02}.

Let us study the situation when $E^2 \gg V\Omega$ and $n \gg 1$.
The former condition is always satisfied for any realistic
situations. Indeed, for a neutron star with $n_n =
10^{38}\thinspace\text{cm}^{-3}$ we get that $V \sim
10\thinspace\text{eV}$. If we suppose that the energy has the
value of several electron-Volts (we will see in Sec.~\ref{BO} that
a bound state can be formed for such low energy neutrinos) and a
neutron star has the angular velocity
$\Omega=10^3\thinspace\text{s}^{-1}(\sim
10^{-13}\thinspace\text{eV})$, we get that the condition $E^2 \gg
V\Omega$ is satisfied. The latter condition, $n \gg 1$, is also
valid (see Sec.~\ref{BO}). It is received there that the critical
value of the quantum number $\mathrm{n}$ at which a bound state is
still possible is equal to $10^{11}$. It means that this condition
is fulfilled.

Taking into account the approximations made above we obtain that
the energy spectrum has the form,
\begin{equation}\label{enlevm}
  E= -V \pm \sqrt{4V\Omega\mathrm{n}+p_z^2+m^2}.
\end{equation}
However, instead of the relation~\eqref{C1C2}, which is valid for
the massless case, we have the corrected one,
\begin{equation}%\label{C1C2}
  \sqrt{4V\Omega {\rm n}}\ C_1+(E+V+4\Omega-p_z)C_2=0.
\end{equation}
In Sec.~\ref{APP} we obtained that the quantity
$p_\mathrm{eff}=\sqrt{4V\Omega\mathrm{n}+p_z^2}$ has the meaning
of the effective momentum of a neutrino. Therefore we get in
Eq.~\eqref{enlevm} that in the limit $\mathrm{n} \gg 1$ the
neutrino energy receives the correction $m^2/2p_\mathrm{eff}$ due
to the non-zero mass, as it should be. Note that this energy
spectrum coincides with the analogous relation found in
Ref.~\cite{BalPopStu09}.

\section{Low energy without flavor mixing: bound states\label{BO}}

In Secs.~\ref{APP} and~\ref{DirEqMASS} we found the basis
spinors~\eqref{etaabs} and the energy levels~\eqref{enlevm} of a
single neutrino mass eigenstate interacting with rotating
background matter. Now we apply these results for the description
of the evolution of the system~\eqref{DireqpsiMP}.

Given the axial symmetry of the problem, we use cylindrical
coordinates and the eigenfunctions
$\psi_a^\mathrm{T}=(\xi_a,\eta_a)$ of Eq.~\eqref{DireqpsiMP} must
be of the form $\eta_a \sim \eta_a(r,\phi) e^{-\mathrm{i}(E_a t -
p_z z)}$. We find the energy eigenvalues in terms of an index ``n"
[see Eq.~\eqref{enlevm}],
\begin{align}\label{enlevMP}
  E_{\rm n}^{(a)\pm{}} = & - V_a \pm \sqrt{4V_a \Omega \mathrm{n} +
  p_z^2+m_a^2},
  \notag
  \\ 
  & {\rm n}=0, 1, 2,\ldots
\end{align}
Here $E_{\rm n}^{(a)+}$ is the energy of a particle (neutrino),
containing an attractive potential ``$-V_a$", while the negative
value $E_{\rm n}^{(a)-}$ must be understood as $-E_{\rm
n}^{(a)-}$, the positive energy of an antiparticle (antineutrino),
containg a potential term ``$+V_a$", which is repulsive. We
included the neutrino mass $m_a$ in the above expression for the
energy, although we will neglect it in the spinors.

In a rotating medium, the effect of rotation is largest for
neutrinos propagating in the equatorial plane. Choosing $Z$ as the
rotation axis, we then look for solutions which are
$z$-independent, i.e. $p_z =0$.

The corresponding two-component spinors are given in terms of
Laguerre functions $I_{{\rm n},s}(\rho)$ with an energy index ``n"
and a radial index ``$s$" [see Eqs.~\eqref{etaabs}
and~\eqref{C1C2}]:
\begin{align}\label{basspinMP}
  u_{a,{\rm n}s}^{(\pm)}(r,\phi) = &
  \sqrt{\frac{V_a\Omega}{2\pi}}
  \begin{pmatrix}
    I_{{\rm n}-1,s}(\rho_a) e^{\mathrm{i}(l-1)\phi} \\
    \mp\,\mathrm{i} I_{{\rm n},s}(\rho_a) e^{\mathrm{i}l\phi} \
  \end{pmatrix},
  \notag
  \\
  & l = {\rm n}-s,
\end{align}
and where $\rho_a = V_a \Omega r^2$, $a=1,2$, is a dimensionless
radial coordinate. For further details of the derivation of
Eqs.~\eqref{enlevMP} and~\eqref{basspinMP} the reader is referred
to Secs.~\ref{APP} and~\ref{DirEqMASS}.

In this case, when ``n" is a non-negative integer, the energies in
Eq.~\eqref{enlevMP} are discrete values and the wavefunctions
decay to zero as $r\to \infty$. This is a required condition for
low enough energies, see discussion below Eq.~\eqref{wfout},
otherwise the wavefunctions would diverge inside the star before
reaching the star radius.

Instead, for neutrinos with larger energies, the wavefunction
reaches the edge of the star and should continue outside. For
those cases the neutrino energy and the index ``n" become
continuous variables (we change the name $\mathrm{n} \to \kappa$
in this case). The solution inside the star now has the more
general form:
\begin{multline}\label{etabigenergy}
  u_{a,\kappa}^{(\mathrm{in})}(r,\phi) = %&
  e^{-\rho_a/2} \rho_a^{l/2}
  \frac{e^{\mathrm{i}l\phi}}{l!}
  \sqrt{\frac{2 V_a \Omega}{2\pi}}
  %\notag
  \\
  \times
  \begin{pmatrix}
    C_1^{(\mathrm{in})} l
    F(l-\kappa,l,\rho_a) e^{-\mathrm{i}\phi}/\sqrt{\kappa \rho_a} \\
    \mathrm{i} C_2^{(\mathrm{in})} 
    F(l-\kappa,l+1,\rho_a) \
  \end{pmatrix},
\end{multline}
where $F(\alpha,\beta,z)$ is a confluent hypergeometric function
and the coefficients $C_{1,2}^{(\mathrm{in})}$ satisfy the
relation
\begin{equation}\label{C1C2cont}
  \sqrt{4 V_a \Omega \kappa}\ C_1^{(\mathrm{in})} +
  (E_a+V_a-p_z)\ C_2^{(\mathrm{in})} = 0,
\end{equation}
considering the energy $E_a$ with the same expression as in
Eq.~\eqref{enlevMP}, but with n$\to\kappa$, a continuous variable.
We derive Eq.~\eqref{C1C2cont} from Eq.~\eqref{C1C2} changing the
norm of the coefficients $C_{1,2}$ there as:
\begin{equation}
  C_{1,2}^{(\mathrm{in})} \to C_{1,2}^{(\mathrm{in})}
  \sqrt{\frac{\mathrm{n}!}{(\mathrm{n}-l)!}},
\end{equation}
and using the property of the confluent hypergeometric function,
$F(l-\mathrm{n},l+1,\rho)=l!Q_{\mathrm{n}-l}^l(\rho)/\mathrm{n}!$.

For the wave function outside the star one must use the outgoing
wave solution in vacuum given in Eq.~\eqref{eta0cylvacrw}, i.e.
\begin{multline}\label{wfout}
  u^{(\mathrm{out})}_a(r,\phi) =
  \frac{1}{\sqrt{2\pi}}
  \\
  \times
  \begin{pmatrix}
    C_1^{(\mathrm{out})} H_{l-1}^{(1)}(p_{\perp{}}r) e^{\mathrm{i}(l-1)\phi} \\
    \mathrm{i} C_2^{(\mathrm{out})} H_{l}^{(1)}(p_{\perp{}}r) e^{\mathrm{i}l\phi} \
  \end{pmatrix},
\end{multline}
where $H_{l}^{(1)}$ are Hankel functions of the first kind and
$p_{\perp{}}= \sqrt{E^2-p_z^2-m_a^2}$ is the momentum
perpendicular to the rotation axis.

The coefficients $C_{1,2}^{(\mathrm{in})}$ and
$C_{1,2}^{(\mathrm{out})}$ are related due to the continuity of
the wave functions~\eqref{etabigenergy} and~\eqref{wfout} at the
neutron star surface,
$u_{a,\kappa}^{(\mathrm{in})}(R,\phi)=u_{a}^{(\mathrm{out})}(R,\phi)$.
Eqs.~\eqref{C1C2cont}, \eqref{c12rel1} and~\eqref{flux} completely
define the coefficients for the solution corresponding to an
unbound wave function.

\begin{figure}
  \centering
  \includegraphics[scale=.4]{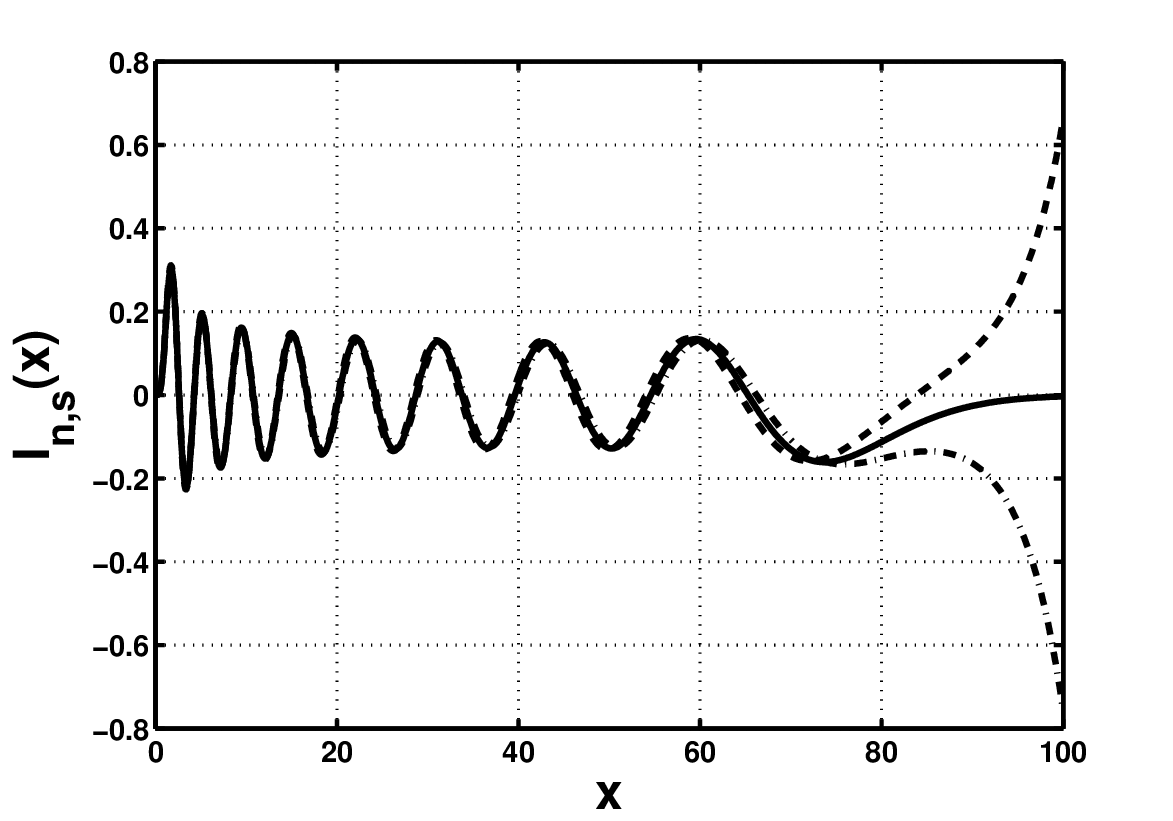}
  \caption{\label{fig2}
  The solutions of Eq.~\eqref{ueq} for $l=10$ and $s=15$ (solid line),
  $l=10$ and $s=15.1$ (dashed line) as well as
  $l=10$ and $s=14.9$ (dash-dot line). The function $I_{\mathrm{n},s}(x)$ at
  non-integer "$s$" should be understood as the confluent hypergeometric function
  $F(\kappa-l,l+1,x)$ in Eq.~\eqref{etabigenergy}.}
\end{figure}

As mentioned above, neutrinos could form bound states inside a
neutron star, provided the energy is small enough. In those cases
the energy \eqref{enlevMP} assumes discrete values, because
otherwise the wave function would diverge before reaching the edge
of the star. In Fig.~\ref{fig2} we present an example of wave
function [see Eq.~\eqref{basspinMP}] for $l=10$ and $s=15$ (solid
line).
It can be seen that the solution is a rapidly oscillating function
in the inner regions and approaches zero for large radii. We also
present in Fig.~\ref{fig2} the wave functions for $l=10$ and
$s=15.1$ (dashed line) as well as for $l=10$ and $s=14.9$
(dash-dot line), showing the divergence at large radii.

From the fact that $E_{\rm n}^{(a)+}$ can be negative for a medium
dominated by neutrons, as seen in Eq.~\eqref{enlevMP}, neutrinos
of low kinetic energy can form bound states inside the star. On
the other hand, antineutrinos cannot be bound because in their
case the potential is repulsive. Limiting the analysis to $p_z=0$
only, the energy spectrum for the bound states is discrete. The
maximum value of ``n" for bound states can be estimated with the
equation $E_{\rm n}^{(a)+}=0$,
\begin{equation}\label{nbound}
  {\rm n} \le \frac{V_a}{4\Omega}.
\end{equation}
This condition is independent from (but consistent with) the
condition that the wave function should not diverge before
reaching the star radius. The position of the last maximum in the
wave function is approximately at $\rho_\mathrm{crit} \sim
2(n+s)$, and this point should be inside the star,
i.e.~$\rho_\mathrm{crit}< V_a \Omega R^2$, where $R$ is the star
radius (remember the definition $\rho = V\Omega r^2$). This
condition means $\mathrm{n} < V_a\Omega R^2/2$.

We can estimate the index ``n" in these bounds for the case of a
neutron star. Taking a reference value for the neutron density
$n_n = 10^{38}\thinspace\text{cm}^{-3}$ (close to nuclear
density), we get a potential $V_a \sim 7$ eV. This is much larger
than the natural neutrino masses expected from oscillation
experiments, so the mass in Eq.~\eqref{enlevMP} and
Eq.~\eqref{nbound} is indeed negligible. Now, for the angular
speed $\Omega$, we can take the Keplerian angular velocity
$\Omega_\mathrm{max}=\sqrt{GM/R^3}$ (see Ref.~\cite{Gho07}) as a
rough upper limit. Taking $M=1.5 M_{\odot{}}$ and
$R=10\thinspace\text{km}$ one obtains for $\Omega_\mathrm{max}\sim
10^4\thinspace\text{s}^{-1}$ (typical values are two or more
orders of magnitude smaller). Eq.~\eqref{nbound} then gives n
$\sim 10^{11}$, a rather large number of bound states as well as
nodes within the star. On the other hand, the condition for the
wave function to fade before reaching the edge, $\mathrm{n} <
V_a\Omega R^2/2$, gives also n $\sim 10^{11}$.

The particles inside the neutron star can be created in the form
of neutrino-antineutrino pairs~\cite{Yak01}. The fact that low
energy neutrinos can be trapped by the star and antineutrinos will
be accelerated out of the star makes it possible the existence of
the low energy antineutrinos luminosity.

The validity of the wave function method has one main drawback
concerning wave coherence: it must be assumed that incoherent
scattering can be neglected all over the star. This is
particularly unlikely if the number of nodes is as large as
$10^{11}$. Another drawback of this treatment concerns neutrino
production: most neutrinos are produced with energies much larger
than a few electron-Volts, and again, the production is in general
an incoherent process.

\section{Effect of rotation on flavor
mixing of low energy neutrinos\label{LO}}

So far we have neglected the off-diagonal elements of
Eq.~\eqref{gab} which mix the neutrino mass states. Let us now
include them as perturbations to study the effect of rotation in
the mixing of neutrino flavors. The general solution of
Eq.~\eqref{DireqpsiMP} can be stated in the following form (see
also Ref.~\cite{approach}):
\begin{align}\label{gensol}
  \eta_a(r,\phi,t)= &
  \sum_{\mathrm{n},s=0}^{\infty}
  \Big(
    a_{{\rm n}s}^{(a)}\  u_{a,{\rm n}s}^{+{}}(r,\phi)
    \exp[-\mathrm{i}E_{\rm n}^{(a)+{}} t]
    \notag
    \\
    & +
    b_{{\rm n}s}^{(a)}\  u_{a,{\rm n}s}^{-{}}(r,\phi)
    \exp[-\mathrm{i}E_{\rm n}^{(a)-{}} t]
  \Big),
  \notag
  \\ &
  a=1,2.
\end{align}
In Eq.~\eqref{gensol} we consider the full state as an expansion
in the eigenstates with time-dependent coefficients $a_{{\rm
n}s}^{(a)}(t)$ and $b_{{\rm n}s}^{(a)}(t)$, requiring the
expansion to satisfy the evolution equation~\eqref{DireqpsiMP},
and imposing the initial condition that the neutrino wavefunctions
at $t=0$ are of flavor $\beta = e$ only.

Substituting the solution into Eq.~\eqref{DireqpsiMP} and using
orthonormality of the spinors in Eq.~\eqref{basspinMP} we obtain
the following set of ordinary differential equations for the
functions $a_{{\rm n}s}^{(a)}(t)$:
\begin{align}\label{adoteq}
  \mathrm{i} \frac{\mathrm{d}}{\mathrm{d}t}{a}_{{\rm n}s}^{(a)}(t)
  = &
  \sum_{{\rm n}',s'=0}^{\infty}
     %\left\{
       \int u_{a,{\rm n}s}^{+{}\dagger}(\mathbf{r}) (g^\mu_{ab}
       \bar\sigma_\mu )
       u_{b,{\rm n}'s'}^{+{}}(\mathbf{r}) \mathrm{d}^2 \mathbf{r}
     %\right\}
     \notag
     \\
     & \times
     \exp[\mathrm{i}(E_{\rm n}^{(a)+{}}-E_{{\rm n}'}^{(b)+{}}) t ]\
     a_{{\rm n}'s'}^{(b)}(t)
     \notag
     \\
     & +
     %\left\{
       \int u_{a,{\rm n}s}^{+{}\dagger}(\mathbf{r}) \Big(g^\mu_{ab}\,\bar\sigma_\mu\Big)
       u_{b,{\rm n}'s'}^{-{}}(\mathbf{r}) \mathrm{d}^2 \mathbf{r}
     %\right\}
     \notag
     \\
     & \times
     \exp[\mathrm{i}(E_{\rm n}^{(a)+{}}-E_{{\rm n}'}^{(b)-{}}) t]\
     b_{{\rm n}'s'}^{(b)}(t),
     \notag
    \\ &
     a \neq b,
\end{align}
and a similar set of equations  for $b_{{\rm n}s}^{(a)}(t)$.

%Let us now apply this formulation to the neutrinos emitted during
%a supernova collapse and formation of a neutron star at the core.
%We will show that the effect of rotation on neutrino emission is
%largest in the few tens of seconds after
%neutronization~\cite{GiuKim07}. At that time, electrons, protons
%and neutrons in the core are degenerate and near beta equilibrium;
%protons and neutrons are non-relativistic whereas electrons are
%ultra-relativistic, so in the ideal Fermi gas approximation their
%densities obey the simple relation:
%%
%\begin{equation}\label{nedeg}
%  n_p = n_e \approx \frac{3\pi^2}{(2m_n)^3}n_n^2,
%\end{equation}
%%
%where $m_n$ is the neutron mass.

Now, since we are interested in the neutron star matter, we can
assume densities close to nuclear matter or above, dominated by
neutrons ($n_e \ll n_n$). Within this approximation, the spinor
products in Eq.~\eqref{adoteq} can be easily calculated,
\begin{align}\label{matrelapart}
  %\notag
  \int & u_{a,{\rm n}s}^{+{}\dagger}(\mathbf{r}) ( g^\mu_{ab}
  \bar\sigma_\mu )
  u_{b,{\rm n}'s'}^{+{}}(\mathbf{r}) \mathrm{d}^2 \mathbf{r}
  \notag
  \\
  & =
  \int u_{a,{\rm n}s}^{-{}\dagger}(\mathbf{r}) ( g^\mu_{ab}
  \bar\sigma_\mu )
  u_{b,{\rm n}'s'}^{-{}}(\mathbf{r}) \mathrm{d}^2 \mathbf{r}
  %&
  %\\
  %\notag%label{matrelpart}
  \notag
  \\
  & \approx
  g^0_{12}\ \delta_{ll'} \delta_{ss'} -
  \frac{g^0_{12}}{2}\sqrt{\frac{\Omega}{V}} \delta_{ll'}
  \notag
  \\ &
  \times
  (2\delta_{ss'}\sqrt{{\rm n}}-\sqrt{s+1}\delta_{s,s'-1}-\sqrt{s}\delta_{s,s'+1}),
  \notag
  \\
  \int & u_{a,{\rm n}s}^{+{}\dagger}(\mathbf{r}) ( g^\mu_{ab}
  \bar\sigma_\mu )
  u_{b,{\rm n}'s'}^{-{}}(\mathbf{r}) \mathrm{d}^2 \mathbf{r}
  \notag
  \\
  & =
  \int u_{a,{\rm n}s}^{-{}\dagger}(\mathbf{r}) ( g^\mu_{ab}
  \bar\sigma_\mu )
  u_{b,{\rm n}'s'}^{+{}}(\mathbf{r}) \mathrm{d}^2 \mathbf{r}
  \notag
  \\
  & \approx
  - \frac{g^0_{12}}{2}\sqrt{\frac{\Omega}{V}} \delta_{ll'}
  \notag
  \\ &
  \times
  (\sqrt{s}\delta_{s,s'+1}-\sqrt{s+1}\delta_{s,s'-1}),
\end{align}
where $l = {\rm n}-s$, $l' = {\rm n}'- s'$, $g_{12}^0$ is the
potential that mixes neutrino types, given in Eq.~\eqref{gab}, and
$V=G_F n_n/\sqrt{2}$ is the approximation of $V_1\approx V_2$
neglecting the electron density $n_e$. Here we must keep terms
linear in $n_e$, otherwise $g_{12}^0$ would vanish, and the case
of mixing would become trivial. To linear order in $n_e$, we can
take the arguments of the Laguerre functions in
Eq.~\eqref{basspinMP} to be equal, $\rho_1 = \rho_2$, neglecting
$n_e$ in Eq.~\eqref{lambda12}. To derive these results we used the
properties \eqref{recurr} and \eqref{integrals} of the Laguerre
functions, given in Eq.~\eqref{integrals}.

The time evolution for ${a}_{{\rm n}s}^{(a)}(t)$ given in
Eq.~\eqref{adoteq}, which is due to the flavor off-diagonal
perturbation, has contributions from many levels $\{ {\rm n},
s\}$, making a general analysis rather complicated. One
simplification occurs if we disregard neutrino-antineutrino
creation or annihilation, which means no transitions between
$u^{+{}}$ and $u^{-{}}$ states. In what follows we disregard
antineutrinos, so we neglect the coefficients $b_{{\rm
n}s}^{(a)}(t)$ altogether. Another simplification arises if we
consider wave packets very narrow in energy, so that only the
nearest states are involved. A final simplification occurs in the
case $s \gg l$, where level transitions are negligible, thus
reducing the problem to solving a two-state quantum system.

%Note the effect, which is analogous to transitions between
%$u^{+{}}$ and $u^{-{}}$ states, induced by the rotating matter,
%was previously noticed in Ref.~\cite{Muk05}. It was shown that the
%generation of asymmetry between neutrinos and antineutrinos is
%possible in axially symmetrical gravitational fields, e.g., in
%Kerr space-time. The electroweak interaction of neutrinos and
%background matter should be formulated in the frame where
%background fermions are at rest or moving with constant velocity.
%Therefore, if we study neutrino interactions with rotating matter,
%we should place an observer in the rotating, i.e. non-inertial,
%frame. It is known that accelerated frames are equivalent to the
%existence of a gravitational field (see, e.g.,
%Ref.~\cite{Wei72p67}). This observation shows that the transitions
%between $u^{+{}}$ and $u^{-{}}$ states in the rotating matter are
%similar to the result of Ref.~\cite{Muk05}.

The situation $s \gg l$, occurs when the neutrino angular momentum
at emission is small, corresponding to wave functions which are
large near the center of the neutron star. In this case the
coefficients $a_{{\rm n}s}^{(a)}$ are, in practice, functions of
just one quantum number $s$, since $s ={\rm n}-l$ while $l$ is
negligible. Assuming $a_{s}^{(a)}$ depend smoothly on $s$, i.e.
$a_{s}^{(a)} \approx a_{s \pm 1}^{(a)}$, and using
Eq.~\eqref{enlevMP}, we get an evolution equation of the form,
\begin{gather}
  \mathrm{i} \frac{\mathrm{d}}{\mathrm{d}t}
  \begin{pmatrix}
   \tilde{a}_{s}^{(1)}\\
   \tilde{a}_{s}^{(2)} \
  \end{pmatrix} =
  \begin{pmatrix}
    \omega/2 & \Delta \\
    \Delta & -\omega/2 \
  \end{pmatrix}
  \begin{pmatrix}
    \tilde{a}_{s}^{(1)}\\
    \tilde{a}_{s}^{(2)} \
  \end{pmatrix},
  \notag
  \\
  \label{psiprimeeq}
  \tilde{a}_{s}^{(1)} = a_{s}^{(1)}e^{-\mathrm{i}\omega t/2},
  \quad
  \tilde{a}_{s}^{(2)} = a_{s}^{(2)}e^{\mathrm{i}\omega t/2},
\end{gather}
where
\begin{align}\label{HamPars}
  \omega = & E_s^{(1)+{}}-E_s^{(2)+{}} =
  V_2-V_1 
  \notag
  \\
  & + \sqrt{4V_1\Omega s +m_1^2} -\sqrt{4V_2\Omega s +m_1^2},
  \notag
  \\
  \Delta = & g^0_{12}.
\end{align}
Since, for non-bound states, $s$ is a very large and continuous
variable, we can define a continuous ``momentum" variable inside
the medium, $p_\mathrm{eff} = \sqrt{4V\Omega s}$, chosen to be the
same for both neutrino types, as it is usually done in the
treatment of non-rotating media~\cite{kmomentum}.  In terms of
more conventional quantities, $\Delta$ and $\omega$ are then
\begin{align}\label{defpartham2}
  \omega = &
  \frac{\delta m^2}{2p_\mathrm{eff}}-
  \sqrt{2} G_\mathrm{F} n_e \cos 2\theta,
  \notag
  \\
  \Delta = & \frac{G_\mathrm{F}}{\sqrt{2}} n_e \sin 2\theta,
\end{align}
In Eqs.~\eqref{HamPars} and~\eqref{defpartham2} we take into
account the non-zero masses of the neutrino mass eigenstates
$m_{1,2}$ and use the standard notation $\delta m^2 =
m_1^2-m_2^2$. As one can see, Eqs.~\eqref{psiprimeeq}
and~\eqref{defpartham2} are practically independent of the
rotation velocity $\Omega$, and so the evolution equation reduces
to the known case of neutrino oscillations in non-moving
background matter.

For the rotation to have any significant effect on neutrino flavor
oscillations the linear velocities of the matter motion should
reach very high values. For the realistic angular velocities of
pulsars $\sim 10^3\thinspace\text{s}^{-1}$ and a neutron star with
radius of $\sim 10-20\thinspace\text{km}$ the matter velocity can
be about $0.1$, which is not enough to significantly change the
transition probability. Therefore the effect potentially
interesting from the phenomenological point of view is the
trapping of low energy neutrinos discussed in Sec.~\ref{BO}.

\section{Conclusion\label{CONCL}}

Summarizing we mention that we studied neutrino flavor
oscillations in rotating matter. The analysis was carried out in
frames of the relativistic quantum mechanics. We used the exact
solutions of the wave equation for a neutrino weakly interacting
with inhomogeniously moving matter to solve the initial condition
problem for the system of mixed flavor neutrinos. Note that the
use of the relativistic quantum mechanics method~\cite{approach}
is essential in describing neutrino flavor oscillations since it
takes into account the coordinate dependence of the neutrino wave
functions~\eqref{basspinMP}. It was possible to derive the
Schr\"{o}dinger like evolution equation for the two component
neutrino wave function for the important case of neutrinos with
small angular momentum. This situation corresponds to the
particles emitted close to the central region of a neutron star.
It was found that rotation does not change the dynamics of
neutrino flavor oscillations, i.e. the quantum mechanical
description of neutrino oscillations is insensitive to the
rotation of background matter.

We have studied the possibility for the existence of neutrino
bound states inside a neutron star. The cases of neutrinos and
antineutrinos are different. We revealed that low energy neutrinos
can be trapped by the rotating neutron star whereas antineutrinos
always escape. The applicability of the relativistic quantum
mechanics method was analyzed. Despite this approach allows one to
account for the interaction with external fields exactly it cannot
be used for the description of the evolution of high energy
neutrinos in dense rotating matter since these particles
experience many incoherent collisions and it is difficult to form
the coherent cylindrical wave inside a star. It means that the
relativistic quantum mechanics is applicable for later stages of
the neutron star evolution when one can neglect the multiple
neutrino collisions with background matter.

\section*{Acknowledgments}
The work has been supported by Conicyt (Chile), Programa Bicentenario PSD-91-2006 and by FAPESP (Brazil). 
The author is thankful to C.~O.~Dib and
J.~Maalampi for helpful discussions.

\appendix

\section{Solution to the wave equation for a neutrino in vacuum
in cylindrical coordinates\label{INICOND}}

In this Appendix we find the solution of the wave equation for a
neutrino in vacuum in cylindrical coordinates. In the chiral basis
for the Dirac $\gamma$-matrices [see Eq.~\eqref{chiralbasis}] the
Dirac field $\psi(x)$ is expressed in terms of two 2-component
chiral fields, $\eta(x)$ and $\xi(x)$ as $\psi^\mathrm{T} =(\xi,
\eta)$. In cylindrical coordinates $(r,\phi,z)$, the spinors $\xi$
and $\eta$ satisfy the differential equations
\begin{align}\label{etacylvac}
  \mathrm{i}\dot{\xi} = & (\bm{\sigma}\cdot\mathbf{p})\xi - m\eta,
  \notag
  \\
  \mathrm{i}\dot{\eta} = & -(\bm{\sigma}\cdot\mathbf{p})\eta - m\xi.
\end{align}
We look for stationary solutions of Eq.~\eqref{etacylvac}, which
must be of the form,
\begin{align}
  \xi(\textbf{r},t) = & e^{-\mathrm{i}(E t - p_z z)} w(r,\phi),
  \notag
  \\
  \eta(\textbf{r},t) = & e^{-\mathrm{i}(E t - p_z z)} u(r,\phi),
\end{align}
where $E$ is the energy and $p_z$ is the $z$ component of the
neutrino momentum. We can separate the variables $(r,\phi)$
introducing radial functions $G_{1,2}(r)$ and $F_{1,2}(r)$,
\begin{align}
  w_1(r,\phi) = & \frac{1}{\sqrt{2\pi}}e^{\mathrm{i} (l-1) \phi} G_1(r),
  \notag
  \\
  w_2(r,\phi) = & \frac{1}{\sqrt{2\pi}}e^{\mathrm{i} l \phi} G_2(r),
  \notag
  \\
  u_1(r,\phi) = & \frac{1}{\sqrt{2\pi}}e^{\mathrm{i} (l-1) \phi} F_1(r),
  \notag
  \\
  u_2(r,\phi) = & \frac{1}{\sqrt{2\pi}}e^{\mathrm{i} l \phi} F_2(r),
\end{align}
where $l$ measures the $z$-component of the orbital angular
momentum. For the radial functions $G_{1,2}$ and $F_{1,2}$ we
obtain the system of the differential equations,
\begin{align}\label{Fcylvac}
  (E - p_z)G_1
  + &
  \mathrm{i}
  \left(
    \partial_r+\frac{l}{r}
  \right)G_2 
  \notag
  \\ &
  = - m F_1,
  \notag
  \\
  (E+p_z)G_2 + & \mathrm{i}
  \left(
    \partial_r-\frac{l-1}{r}
  \right)G_1 
  \notag
  \\ &  
  = - m F_2,
  \notag
  \\
  (E+p_z)F_1 - & \mathrm{i}
  \left(
    \partial_r+\frac{l}{r}
  \right)F_2 
  \notag
  \\ &  
  = - m G_1,
  \notag
  \\
  (E-p_z)F_2 - & \mathrm{i}
  \left(
    \partial_r-\frac{l-1}{r}
  \right)F_1 
  \notag
  \\ &
  = - m G_2.
\end{align}
The solutions of Eq.~\eqref{Fcylvac} can be given in terms of
Bessel functions. For standing waves that are regular at the
origin, the functions $w(r,\phi)$ and $u(r,\phi)$ are
\begin{align}\label{eta0cylvac}
  w(r,\phi) & =
  \frac{1}{\sqrt{2\pi}}
  \begin{pmatrix}
    B_1 J_{l-1}(p_{\perp{}}r) e^{\mathrm{i}(l-1)\phi} \\
    \mathrm{i} B_2 J_{l}(p_{\perp{}}r) e^{\mathrm{i}l\phi} \
  \end{pmatrix},
  \notag
  \\
  u(r,\phi) & =
  \frac{1}{\sqrt{2\pi}}
  \begin{pmatrix}
    C_1 J_{l-1}(p_{\perp{}}r) e^{\mathrm{i}(l-1)\phi} \\
    \mathrm{i} C_2 J_{l}(p_{\perp{}}r) e^{\mathrm{i}l\phi} \
  \end{pmatrix},
\end{align}
where $p_{\perp{}}^2 = E^2 - p_z^2-m^2$, $B_{1,2}$ and $C_{1,2}$
are the undefined coefficients.

Using the identity for the Bessel functions,
\begin{align}
  J_l'(x) + & \frac{l}{x}J_l(x)=J_{l-1}(x),
  \notag
  \\
  J_{l-1}'(x) - &\frac{l-1}{x}J_{l-1}(x)=-J_l(x),
\end{align}
and Eq.~\eqref{Fcylvac} we find that the coefficients $B_{1,2}$
and $C_{1,2}$ obey the system of the following algebraic
equations:
\begin{align}%\label{Fcylvac}
  (E-p_z)B_1-p_{\perp{}}B_2= & -m C_1,
  \notag
  \\
  (E+p_z)B_2-p_{\perp{}}B_1= & -m C_2,
  \notag
  \\
  (E+p_z)C_1+p_{\perp{}}C_2= & -m B_1,
  \notag
  \\
  (E-p_z)C_2+p_{\perp{}}C_1= & -m B_2.
\end{align}
In the limit of the small neutrino mass one can see that the
equations for $B_{1,2}$ and $C_{1,2}$ decouple giving one relation
only for the coefficients $C_{1,2}$,
\begin{equation}\label{c12rel1}
  \sqrt{E+p_z}C_1+\sqrt{E-p_z}C_2=0.
\end{equation}

Normalizing the wave function as
\begin{multline}\label{wfnormcyl1}
  \int \mathrm{d}^2\mathbf{r} \thinspace
  u_{E,\, l}(r,\phi)^\dag u_{E',\, l'}(r,\phi)
  \\ =
  E\, \delta(E-E') \, \delta_{ll'},
\end{multline}
and using the integral property of Bessel functions
\begin{equation}
  \int_0^{\infty} \mathrm{d}r \thinspace
   r J_l(kr)J_l(k'r)=\frac{1}{k}\delta(k-k'),
\end{equation}
one finds that the coefficients $C_{1,2}$ also must satisfy the
relation
\begin{equation}\label{c12rel2}
  C_1^2+C_2^2=E^2.
\end{equation}

Instead of standing waves like Eq.~\eqref{eta0cylvac}, we can have
outgoing waves. These solutions are similar, but in terms of
Hankel functions of the first kind, $H_{l}^{(1)}(x)$,
\begin{equation}\label{eta0cylvacrw}
  u(r,\phi) =
  \frac{1}{\sqrt{2\pi}}
  \begin{pmatrix}
    C_1 H_{l-1}^{(1)}(p_{\perp{}}r) e^{\mathrm{i}(l-1)\phi} \\
    \mathrm{i} C_2 H_{l}^{(1)}(p_{\perp{}}r) e^{\mathrm{i}l\phi} \
  \end{pmatrix}.
\end{equation}
The Hankel functions have the asymptotic behaviour of outgoing
radial waves in the two dimensional plane
\begin{equation}
  H_{l}^{(1)}(x) \approx \sqrt{\frac{2}{\pi x}}\exp
  \left[
    \mathrm{i}
    \left(
      x - \frac{\pi l}{2} - \frac{\pi}{4}
    \right)
  \right].
\end{equation}
Instead of the normalization~\eqref{wfnormcyl1} one has to define
the particle flux at large distances, $j_\infty$. For a radial
flow in two dimensions, the flux of particles through a circle of
the large radius $r$ has the form,
\begin{equation}%\label{flux}
j_\infty = \lim_{r\to\infty}  \int_0^{2\pi} r\mathrm{d}\phi\
u^\dag(r,\phi)\,
  u(r,\phi),
\end{equation}
where we have assumed relativistic particles. From this
expression, one can derive norm for the coefficients $C_{1,2}$,
\begin{equation}\label{flux}
  C_1^2+C_2^2=\frac{\pi E\, j_\infty}{2}.
\end{equation}
On the other hand, the relation~\eqref{c12rel1} is still valid for
the wave function~\eqref{eta0cylvacrw}.

\bibliographystyle{plain}

\begin{thebibliography}{40}

\bibitem{GiuKim07}
  Giunti, C. and Kim, C.~W.,
  \textit{Fundamentals of neutrino physics and
  astrophysics},
  Oxford University Press, NY, 2007, pp.~511--539.

\bibitem{Yak01}
  Yakovlev, D.~G., Kaminker, A.~D.,  Gnedin, O.~Y. and Haensel, P.,
  Neutrino emission from neutron stars,
  \textit{Phys. Rep.} \textbf{354} 1, 2001, astro-ph/0012122.

\bibitem{KusSer96}
  Kusenko, A. and Segre, G.,
  Velocities of pulsars and neutrino oscillations,
  \textit{Phys. Rev. Lett.} \textbf{77} 4872, 1996, hep-ph/9606428.

\bibitem{sd}
  Mikaelian, K.~O.,
  New mechanism for slowing down the rotation of dense stars,
  \textit{Astrophys. J.} \textbf{214} L22, 1977;
  Epstein, R.,
  Neutrino angular momentum loss in rotating stars,
  \textit{Astrophys. J.} \textbf{219} L39, 1978;
  Dvornikov, M. and Dib, C.~O.,
  The effect of rotation in the neutrino emission from a neutron star,
  \textit{Phys. Rev. D} \textbf{82} 043006, 2010,
  arXiv:0907.1445~[astro-ph].

\bibitem{RinWon04}
  Ringwald, A. and Wong, Y.~Y.~Y.,
  Gravitational clustering of relic neutrinos and implications for their detection,
  \textit{JCAP} \textbf{0412} 005, 2004, hep-ph/0408241.

\bibitem{superf}
  Kapusta, J.,
  Neutrino superfluidity,
  \textit{Phys. Rev. Lett.} \textbf{93} 251801, 2004, hep-th/0407164;
  Bhatt, J.~R. and Sarkar U.,
  Majorana neutrino superfluidity and stability of neutrino dark energy,
  \textit{Phys. Rev. D} \textbf{80} 045016, 2009, arXiv:0805.2482~[hep-ph].

\bibitem{Dvo06}
  Dvornikov, M.,
  Neutrino spin oscillations in gravitational fields,
  \textit{Int. J. Mod. Phys. D}  \textbf{15} 1017, 2006, hep-ph/0601095.

\bibitem{Tha78}
  Guha Thakurta, S.~N.,
  Neutrino trapping in rotating matter.
  \textit{J. Phys. A} \textbf{11} 2213, 1978.

\bibitem{nuinsideNS}
  Loeb, A.,
  Bound neutrino sphere and spontaneous neutrino pair creation in cold neutron
  stars,
  \textit{Phys. Rev. Lett.} \textbf{64} 115, 1990;
  Kiers, K. and Weiss, N.,
  Coherent neutrino interactions in a dense medium,
  \textit{Phys. Rev. D} \textbf{56} 5776, 1997, hep-ph/9704346.

\bibitem{StuROT}
  Grigoriev, A.~V., Savochkin, A.~M. and Studenikin, A.~I.,
  Quantum states of the neutrino in a nonuniformly moving medium,
  \textit{Russ. Phys. J.} \textbf{50} 845, 2007;
  Studenikin, A.~I.,
  Method of wave equations exact solutions in studies of neutrinos
  and electrons interaction in dense matter,
  \textit{J. Phys. A} \textbf{41} 164047, 2008, arXiv:0804.1417~[hep-ph].

\bibitem{approach}
  Dvornikov, M.,
  Evolution of coupled classical fields,
  \textit{Phys. Lett. B} \textbf{610} 262, 2005,
  hep-ph/0411101;
  Dvornikov, M.,
  Evolution of coupled fermions under the influence of an external axial-vector
  field,
  \textit{Eur. Phys. J. C} \textbf{47} 437, 2006,
  hep-ph/0601156;
  Dvornikov, M.,
  Neutrino oscillations in matter and in twisting magnetic
  fields,
  \textit{J. Phys. G} \textbf{35} 025003, 2008, 
  arXiv:0708.2328~[hep-ph];
  Dvornikov, M.,
  Neutrino flavor oscillations in background matter,
  \textit{J. Phys. Conf. Ser.} \textbf{110} 082005, 2008, 
  arXiv:0708.2975~[hep-ph];
  Dvornikov, M.,
  Evolution of mixed particles interacting with classical
  sources,
  \textit{Phys. Atom. Nucl.} \textbf{72} 116, 2009,
  hep-ph/0610047;
  Dvornikov, M. and Maalampi, J.,
  Evolution of mixed Dirac particles interacting with an external magnetic
  field,
  \textit{Phys. Lett. B} \textbf{657} 217, 2007,
  hep-ph/0701209;
  Dvornikov, M. and Maalampi, J.,
  Oscillations of Dirac and Majorana neutrinos in matter and a magnetic
  field,
  \textit{Phys. Rev. D} \textbf{79} 113015, 2009,
  arXiv:0809.0963~[hep-ph].

\bibitem{BalPopStu09}
  Balantsev, I., Popov, Yu. and Studenikin, A.,
  Neutrino magnetic moment and neutrino energy quantization in rotating
  media,
  \textit{Nuovo Cim. C} \textbf{32} 53, 2009,
  arXiv:0906.2391~[hep-ph].

\bibitem{MohPal04}
  Mohapatra, R.~N. and Pal, P.~B.,
  \textit{Massive neutrinos in physics and astrophysics},
  World Scientific, Singapore, 2004, pp.~96--101.

\bibitem{DvoStu02JHEP}
  Dvornikov, M. and Studenikin, A.,
  Neutrino spin evolution in presence of general external fields,
  \textit{JHEP} \textbf{09} 016, 2002,
  hep-ph/0202113.

\bibitem{LobStu01}
  Lobanov, A.~E. and Studenikin, A.~I.,
  Neutrino oscillations in moving and polarized matter
  under the influence of electromagnetic fields,
  \textit{Phys. Lett. B} \textbf{515} 94, 2001, hep-ph/0106101.

\bibitem{ItzZub80}
  Itzykson, C. and Zuber, J.~B.,
  \textit{Quantum field theory},
  McGraw-Hill, NY, 1980.

\bibitem{SokTer74}
  Sokolov, A.~A. and Ternov, I.~M.,
  \textit{Relativistic electron},
  Nauka, Moscow, 1974, pp.~263--277.

\bibitem{SokTer74p282}
  See p.~282 in Ref.~\cite{SokTer74}.

\bibitem{Bor02}
  Bagrov, V.~G.,
  Quantum theory of relativistic particle radiation,
  in \textit{Radiation theory of relativistic particles}
  (V.~A~Bordovitsyn ed.),
  Moscow, Fizmatlit, 2002, p.~168.

\bibitem{Gho07}
  Ghosh, P.
  \textit{Rotation and acretion powered pulsars},
  World Scientific, Singapore, 2007, pp.~212--214.

\bibitem{kmomentum}
  The arbitrary choice of having the same momentum (and different
  energies), or viceversa, for a coherent beam of two or more
  neutrino types, is justified because it gives the same result as
  the more rigorous treatment of using wave packets with a small
  spread in momentum (and energy), in the case of relativistic
  particles with very small masses.

\end{thebibliography}

\end{document}